\documentclass[12pt]{article}
\usepackage [cp1251]{inputenc}
\usepackage [russian]{babel}
\usepackage{graphicx}
\usepackage {amsmath}
\usepackage {amssymb}
\usepackage {longtable}
\usepackage {multirow}

\title{Surface tension of monoatomic liquids after relaxation: the main energy contributions}
\author{$^{1}$\textbf{Dmitry M. Naplekov},$^{1,2}$\textbf{Vladimir V. Yanovsky}}

\date{}

\begin{document}

\maketitle
$^{1}$\textit{Institute for Single Crystals, NAS Ukraine, 60 Nauky Ave., Kharkov, 61001, Ukraine}

$^{2}$\textit{V. N. Karazin Kharkiv National University, 4 Svobody Sq., Kharkov, 61022, Ukraine}

\begin{abstract}
We consider the atomistic origin and the main mechanisms determining the energy of a liquid interface after relaxation. A simple theory is constructed for the monatomic densely packed liquids that allows calculation of the surface tension coefficients based on the interatomic potential parameters, without fitting coefficients. Considered are the potential energies of both <<broken>> and stretched bonds between surface atoms. The later contribution is found to be from $20 \%$ to $45 \%$, with an average stretching of the first layer in the range of $7 \% - 14 \%$. The equality of the unit tension force and the unit surface energy is shown. The calculated surface tension coefficients relate to the experimental values as $1.0 - 1.3$ for the ten $fcc$ substances considered. For twenty $bcc$ substances, this range is $0.65 - 1.35$. The increase of density of the second atomic layer is predicted, due to its influence on the coordination number of surface atoms. For a convex surface, increase in the surface tension coefficient with increase in curvature is predicted.
\end{abstract}

\section{Introduction.}

Surface tension plays an important role in a number of physical phenomena and processes, such as nucleation, coagulation, sintering, various capillary effects, flows through a porous medium, etc. At a high fraction of surface atoms, as is the case for nano-sized objects, the properties of the entire system are largely determined by its surface. At the same time, the origin and properties of the surface energy are still not clear enough \cite{surf-Zhao}. In particular, the open question is whether the value of the surface tension coefficient $\gamma$ increases or decreases with the radius of curvature of the surface \cite{surf-Bru,surf-Wang,surf-Kim,surf-Duan,surf-Hole}.

The attempts to clarify the origin of the surface tension \cite{surf-Dur,surf-Berr,surf-Hass} and to construct a quantitative theory \cite{surf-Lenn,surf-Zhao,surf-Sop,surf-Dav,surf-Mona} are currently ongoing. Also the phenomenological correlations are discussed between the tension coefficient $\gamma$ and various characteristics of substances \cite{surf-Schy,surf-Weir,surf-Was}, including viscosity \cite{surf-Iida}, speed of sound \cite{surf-Bla}, etc. They give correct in the order of magnitude values of $\gamma$, but do not clarify the nature of this energy. Such correlations arise because physical constants of substances are not independent, but are determined by parameters of the interaction potential.

In this paper, we consider the surface energy to be the potential energy of the bonds of atoms of the top surface layers. Some of these bonds are <<broken>>, others on average, are significantly stretched. The presence of the latter actually leads to the tension of the surface. Along with the stretching of the first layer, the compaction of the second layer is predicted. Such an increase in the density of the transition surface layer was observed in some simulations \cite{surf-Chen}. It is related to the dependence of the number of <<broken>> bonds on the concentration of atoms in the layers. All these are the most influential factors that determine the surface tension coefficient after relaxation.

There are a number of less important factors, in particular those associated with the thermal motion of atoms. They were not taken into account in this paper, since in this case it is necessary to take into account all contributions that change the surface energy by single percents, and since the known experimental values of the tension coefficients differ by about $10 \%$ according to different data \cite{surf-Keen,surf-Nogi}. One of the reasons for the latter is the strong influence of impurities of surface-active substances, such as oxygen, sulfur, etc. Therefore, in this work we consider only the main mechanisms in the simplest case of densely packed monatomic liquids.

\section{Basic model of the surface relaxation.}

\begin{figure}
 \centering
 \includegraphics[width=7 cm]{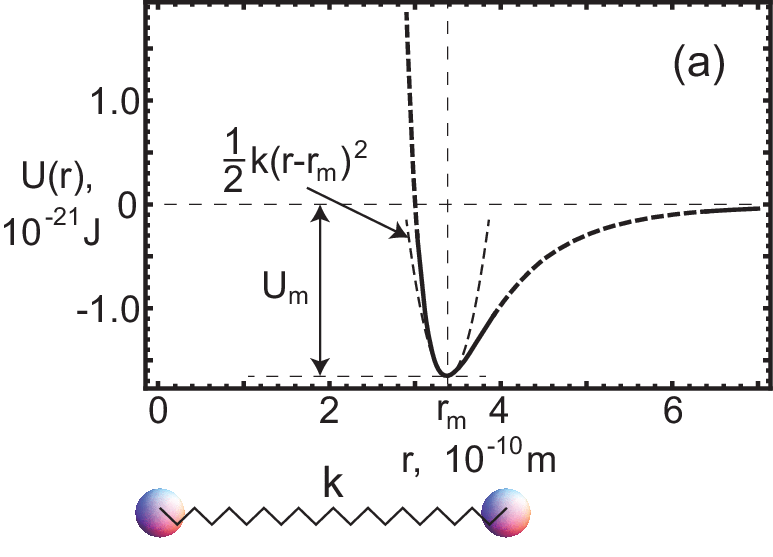}
 \includegraphics[width=6 cm]{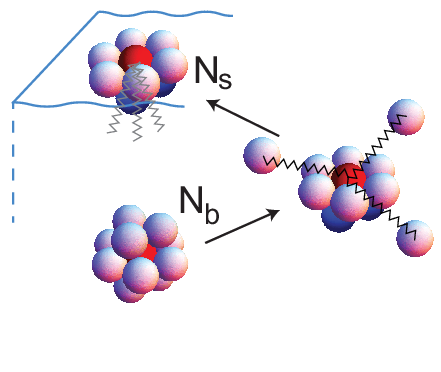}\\
 \caption{(a) Typical atomic interaction potential and its approximation $U(r)=-U_m+\frac{1}{2}k(r-r_m)^2$ in the vicinity of the minimum $r_m$. The atoms may be thought of as connected with a nonlinear spring of stiffness $k$. (b) Simple $fcc$ packing of atoms, neighbours from the first coordination shell are shown. Bringing one bulk atom to the surface requires the extreme stretching or <<breakage>> of $N_b-N_s$ bonds. For this reason, the creation of a new surface requires energy.}
 \label{surf-fig1}
\end{figure}

Let us consider a simple model of a condensed matter interface on an atomic level. We will consider a flat interface between liquid and vacuum or a dilute gas. The liquid will be monoatomic, consisting of identical, densely packed and interacting atoms. The interaction between atoms will be described with a pair potential $U(r)$, its typical form is shown in Fig.\ref{surf-fig1}(a). In the case of a general position, this potential has a minimum at a distance $r_m$, and can be approximately represented as $U(r)=-U_m+\frac{1}{2}k(r-r_m)^2$ in the vicinity of $r_m$. The energy $U_m$ is usually called the energy needed to <<break>> the bond. We will also call it the energy required to ultimately stretch the bond, since strictly speaking the atoms interact at any distance and the potential energy of the bond does not disappear when the bond is <<broken>>. The interatomic interaction can be thought of as a nonlinear spring. If to count the energy from the potential minimum, the energy of the spring will be $\frac{1}{2}k(r-r_m)^2$, while the deviation from the equilibrium position is small $r-r_m \ll r_m$. With an ultimate stretch $r \gg r_m$, its energy approaches a constant value $U_m$ and, due to nonlinearity, practically does not increase with distance any more. Thus, we will use a simplified description of the interaction between atoms, without accepting any exact form of the pair interaction potential. This description is characterized by three parameters: the distance $r_m$, which corresponds to an unstretched bond, the bond rigidity $k$ in the vicinity of this equilibrium position, and the energy of the ultimate stretch or <<break>> of the bond $U_m$.

In the liquid state, the packing of atoms is characterized by some close order \cite{surf-Egam}. We will consider the simplest case of $fcc$ packing, with the bulk atoms having $N_b=12$ nearest neighbors and the surface atoms having a lower number $N_s$ of nearest neighbors. For a non-optimized or non-relaxed atomic structure, it is equal to $N^{no}_s=9$. These numbers of atoms form the first coordination shell for a selected atom. A surface atom and its environment are conventionally shown in Fig.\ref{surf-fig1}(b). In order to bring an atom out of the bulk to the surface of the liquid, it is necessary to put away $N_b-N_s$ of its neighbors. To do this, the corresponding number of bonds must be ultimately stretched, which requires energy $(N_b-N_s) U_m$. This is the core reason why the creation of a new surface requires energy. When a surface atom returns to the bulk, additional atoms will approach it. In this case, the stored in ultimately stretched bonds potential energy will be converted back to the kinetic energy of these atoms. Thus, the surface atoms have more extremely stretched bonds compared to the bulk atoms, which makes their potential energy higher.

Since there is an additional energy associated with the surface atoms, the system will tend to minimize it. Such a tendency is a well-established, general principle. Therefore, the surface will not retain an unrelaxed bulk structure but transfer to some more energetically favorable state in the process of relaxation. The minimization of the surface energy is possible in three ways: by minimizing the total surface area, by minimizing the number of surface atoms in a unit of the remaining area, and by minimizing the energy of each individual atom remaining on the surface. We will consider the entire surface area $S$ fixed. The second and third are the main mechanisms that will be considered below to calculate the surface characteristics after relaxation.

Removal of the atoms from the first surface layer, without changing the surface area, leads to the bonds between the remaining atoms being stretched. In the first approximation, the most energetically favorable state will be determined by the balance of the surface energy decrease due to the decrease in the number of its atoms and the energy increase due to the accompanying stretching of the bonds between the remaining atoms. Such stretching of the bonds between first layer atoms actually leads to the emergence of the tension of the surface.

Since the concentrations of atoms in the first two surface layers are different, we will use the relative concentrations ratio $n$, i.e., the number of atoms of the second layer per one atom of the first layer. Let us also note that stretching of the first atomic layer already changes the numbers of nearest neighbors for the first and second layers atoms, making $N_s$ not equal to $N^{no}_s$.

\begin{figure}
 \centering
 \includegraphics[width=6 cm]{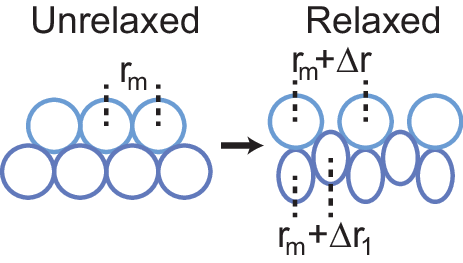}\\
 \caption{The first two layers of the surface atoms are shown. In the relaxed state, the first layer is stretched and the second layer is compressed. The stretched state of the first layer reduces the surface energy by decreasing the number of high-energy atoms per unit area. It also changes the coordination numbers for all the remaining atoms. Compression of the second layer increases the number of nearest neighbors for the first layer atoms, which additionally reduces the surface energy.}
 \label{surf-fig3}
\end{figure}

When further reduction in the number of atoms per unit surface area becomes energetically unfavorable, it is still possible to reduce energy $(N_b-N_s) U_m$ of a single surface atom. This is possible via an increase of the coordination number $N_s$ of each first layer atom. Its nearest neighbors are partially neighboring atoms from the first surface layer $N_{fl}=6$, but the increase of their number is energetically unfavorable. There remain $N_s-N_{fl}$ neighboring atoms from the second layer. The higher the concentration of atoms in the second layer, the greater will be their number per atom of the first layer. Therefore, if the second layer is additionally compacted, the question of finding the most energetically favorable state arises again. On the one hand, compaction means that the bonds between the atoms of the second layer must be compressed, which adds to the surface energy. On the other hand, an increase in the number of atoms of the second layer will increase the number of nearest neighbors for the atoms of the first layer, which decreases the surface energy.

As a result, within the framework of such basic consideration, the most energetically favorable state will be characterized by some average stretches $\Delta r$ and $\Delta r_1$ of the bonds between the atoms of the first and second layers (see Fig.\ref{surf-fig3}). The corresponding relative concentration of atoms $n$ will be:

\begin{equation}\label{surf-eq6}
n=\frac{(r_m+\Delta r)^2}{(r_m+\Delta r_1)^2}
\end{equation}

The surface tension coefficient $\gamma^{ener}$ as the energy of a unit surface area we write as:

\begin{equation}\label{surf-eq5}
\gamma^{ener} = \frac{(N_b-N_s(n)) U_m + \frac{N_{fl}}{2} \frac{k \Delta r^2}{2} + n \frac{N_{fl}}{2} \frac{k \Delta r_1^2}{2} }{\frac{\sqrt{3}}{2} (r_m+\Delta r)^2}
\end{equation}

Here the denominator is the area per one first layer atom, provided that surface has a hexagonal structure. The numerator is the energy of the first and second layers atoms per this area. For small deviations of $n$ from unity, the number of nearest neighbors for a first layer atom $N_s(n)$ can be approximated by a linear dependence on $n$. It should pass through the point $N_s(1)=N^{no}_s$. Considering that it also approximately passes through $N_s(0)=N_{fl}$, we will obtain:

\begin{equation}\label{surf-eq5-1}
N_s(n)=N_{fl}+(N^{no}_s-N_{fl})n
\end{equation}

Both layers in this basic approach are considered to retain a simple hexagonal structure, but of different scales. The stretches of the interlayer bonds are considered to be insignificant. It is clear that all these are approximations, they will be considered further in more detail.

The expression Eq.\ref{surf-eq5} can be minimized by two parameters $\Delta r$ and $\Delta r_1$, which results in the surface tension coefficient:

\begin{equation}\label{surf-eq7}
\gamma^{ener} = \frac{2 N_{fl}}{\sqrt{3}} \frac{N_{fl} (N_b-N^{no}_s) \frac{r_m^2 k}{U_m} - 8 (N_b-N_{fl})(N^{no}_s-N_{fl})}{(N_{fl} \frac{r_m^2 k}{U_m}+4(N_b-N_{fl}))(N_{fl} \frac{r_m^2 k}{U_m}-4(N^{no}_s-N_{fl}))} k
\end{equation}

The relative bond extensions, established in the first and second atomic layers, are:

\begin{equation}\label{surf-eq8}
\begin{array}{l}
\: \frac{\Delta r}{r_m}=4 \: \frac{N_b-N_{fl}}{N_{fl}} \: \frac{U_m}{r_m^2 k}\\
\frac{\Delta r_1}{r_m}=- 4 \: \frac{N^{no}_s-N_{fl}}{N_{fl}} \: \frac{U_m}{r_m^2 k}\\
\end{array}
\end{equation}

Since $N_b > N^{no}_s$, the second surface layer is compressed less than the first layer stretched. Therefore, the surface as a whole remains stretched. For the coefficient $\gamma^{tens}$ as the tension force per unit length, one can verify that (for $N_{fl}=6$):

\begin{equation}\label{surf-eq9}
\gamma^{tens} = \frac{\frac{3}{2} k \Delta r}{\frac{\sqrt{3}}{2}(r_m+\Delta r)}+\frac{\frac{3}{2} k \Delta r_1}{\frac{\sqrt{3}}{2}(r_m+\Delta r_1)} = \gamma^{ener}
\end{equation}

Thus, the minimum of the surface energy corresponds to the equality of coefficients $\gamma^{ener}$ and $\gamma^{tens}$. The coefficients $\frac{3}{2}$ and $\frac{\sqrt{3}}{2}$ in the average force and average length per atom correspond to the random orientation of the system of bonds of the surface atom. The derivation of these coefficients is provided in Appendix 1.

\section{Comparison with experiment.}

To compare the obtained surface tension coefficient Eq.\ref{surf-eq7} with the experimental values, it is necessary to know the parameters of the interaction potential $U_m$, $r_m$ and $k$ for a given substance. We will relate these parameters to other physical properties of the substance, choosing those that are most directly related to a given parameter and most reliably established experimentally.

It is not difficult to estimate the parameter $r_m$ based on the molar volume of the substance $V_m$:

\begin{equation}\label{surf-eq11}
r_m = \sqrt[3]{\frac{6 k_{pac}}{\pi} \frac{V_m}{N_a}}\\
\end{equation}

where $k_{pac}$ is the packing density of atoms, which for the $fcc$ atomic structure is $k_{pac}^{fcc} = \frac{\pi}{3 \sqrt{2}} \approx 0.74$, $N_a$ is the Avogadro number.

The parameter $U_m$ of a substance can be easily related to the isothermal molar heat of vaporization $H_{\nu}$. Evaporation of one atom is associated with the <<breaking>> of all $N_b$ bonds, since the evaporated surface atom is replaced by a new atom from the bulk. It requires the energy $N_b U_m$, which should be equal to $\frac{H_{\nu}}{N_a}$, therefore:

\begin{equation}\label{surf-eq10}
U_m = \frac{H_{\nu}}{N_b N_a}
\end{equation}

The parameter $k$ can be found from the coefficient of isothermal volumetric compressibility of the substance $k_T=\frac{1}{P} \frac{\Delta V}{V}$. This characteristic shows how much the volume of the substance changes when uniform pressure is applied from all sides. The external pressure is compensated by the internal elasticity of the compressed bonds between the atoms. The relationship between the compressibility $k_T$ and the rigidity of one bond $k$ is considered in Appendix 1, where the following expression is derived:

\begin{equation}\label{surf-eq12}
k = \frac{\pi}{4 k_{pac}} \frac{r_m}{k_T}
\end{equation}

\begin{figure}
 \centering
 \includegraphics[width=6 cm]{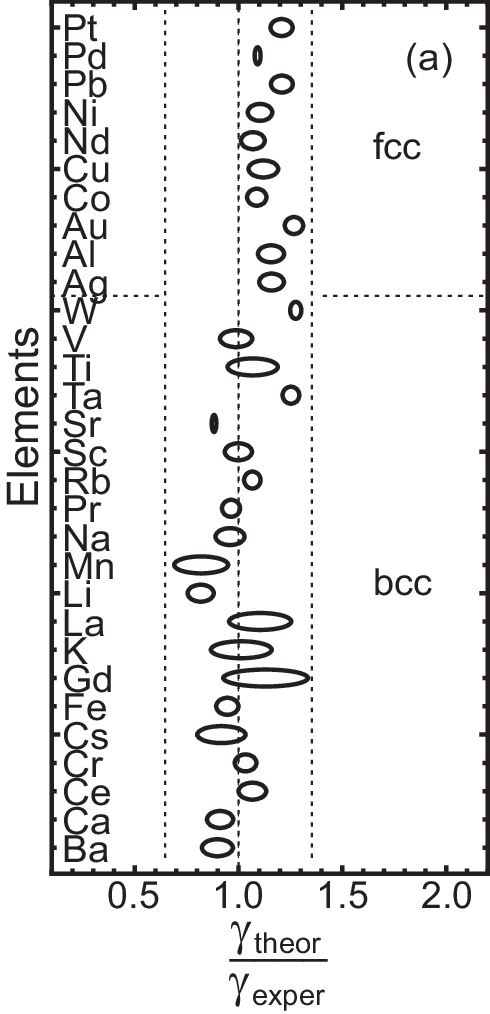}
 \includegraphics[width=6 cm]{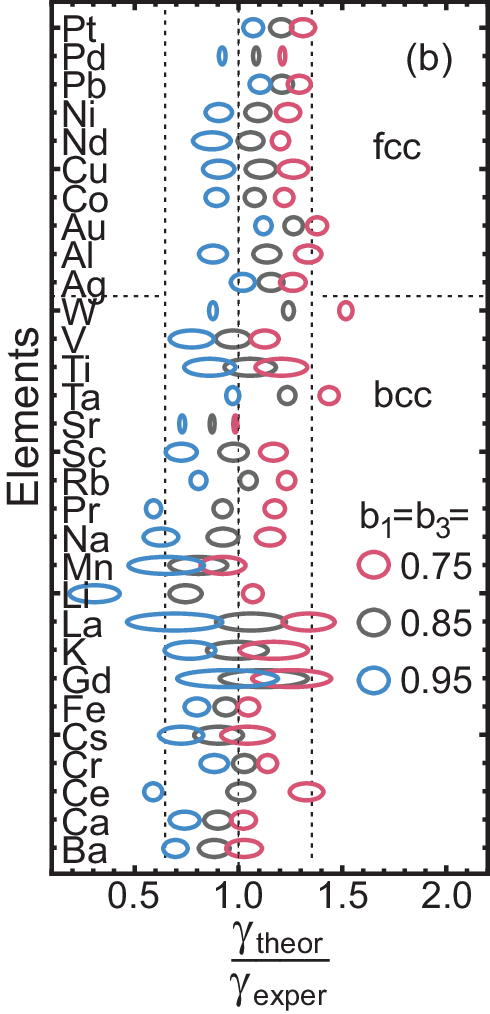}\\
 \caption{The ratio of the theoretical and experimental values of the surface tension coefficients for different substances. The size of the ellipse reflects the maximum range of possible ratio values, due to different experimental data from different authors. (a) According to Eq.\ref{surf-eq7}, without fitting coefficients. (b) According to Eq.\ref{surf-eq9-7} with coefficients $b_1$ and $b_3$ set equal to $(0.75,0.85,0.95)$.}
 \label{surf-fig4}
\end{figure}

Substituting the above expressions into Eq.\ref{surf-eq7}, for the relation of surface tension with $H_{\nu}$, $V_m$ и $k_T$ we receive:

\begin{equation}\label{surf-eq13}
\gamma =N_{fl} \sqrt[3]{\frac{2 \pi^2}{\sqrt{3}k_{pac}^2}}\frac{3 N_b N_{fl} (N_b-N_s) \frac{V_m}{H_{\nu}} - 16 (N_b-N_{fl})(N_s-N_{fl}) k_T}{(3 N_b N_{fl} \frac{V_m}{H_{\nu}} + 8 (N_b-N_{fl}) k_T )(3 N_b N_{fl} \frac{V_m}{H_{\nu}} - 8 (N_s-N_{fl}) k_T)} \, \sqrt[3]{\frac{V_m}{N_a}}
\end{equation}

Since the expression Eq.\ref{surf-eq13} was obtained for the substances with $fcc$ structure, we will compare it with the experimental data for substances with $fcc$ structure at the melting temperature. For other substances, with the $bcc$ structure for example, a similar theory can be constructed with additional accounting for the initial stretching and compression of bonds of the bulk atoms. However, such calculations require more information on the interaction potential, since in this case the bond stretching may not be considered small. In this paper, we will limit ourselves to presenting the results for $fcc$ substances, and also the results of substitution of $bcc$ substances data into the formulas for the $fcc$ structure, since that also leads to a reasonable agreement with the experimental results.

A comparison of the surface tension coefficients obtained using Eq.\ref{surf-eq7} with the experimental values is shown in Fig.\ref{surf-fig4}(a). For all ten $fcc$ substances, the extreme deviations of the ratio of the theoretical and experimental $\gamma$ values are within the range of $1.0 - 1.3$, which can be considered a good agreement for such a simplified theory. The experimental values of $\gamma$ and $k_T$ are known with the same order of accuracy. All values of constants were taken for substances in the liquid state at the melting temperature, the used data are gathered in Appendix 2.

\begin{figure}
 \centering
 \includegraphics[width=8 cm]{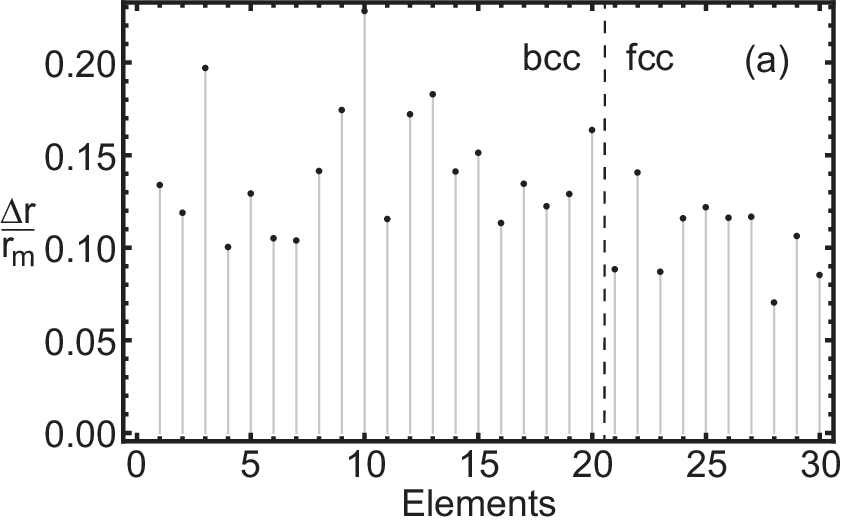}
 \includegraphics[width=8 cm]{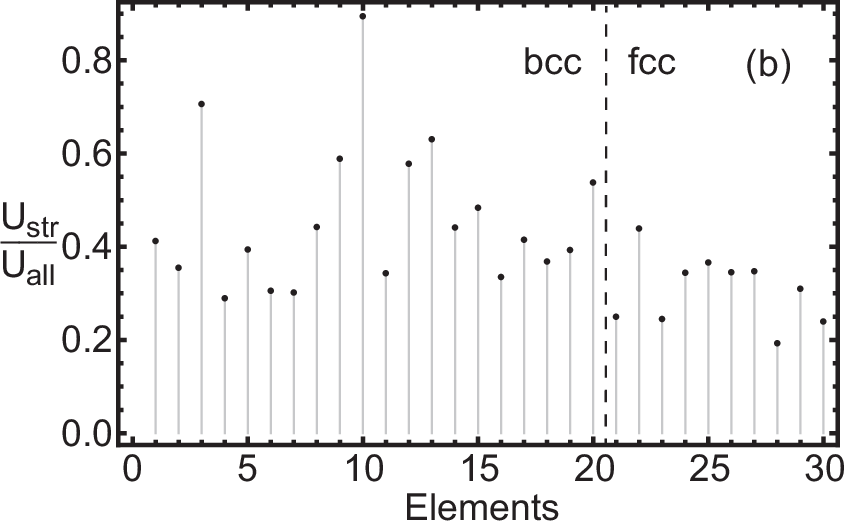}\\
 \includegraphics[width=8 cm]{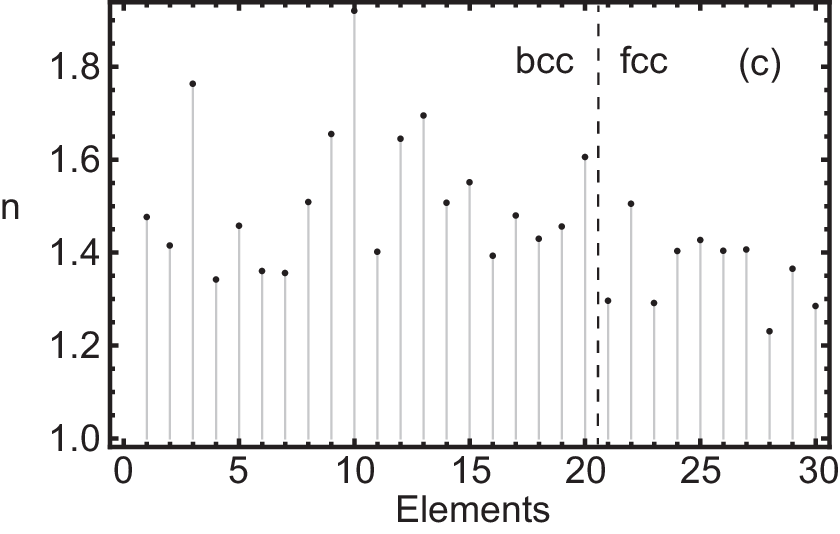}
 \includegraphics[width=8 cm]{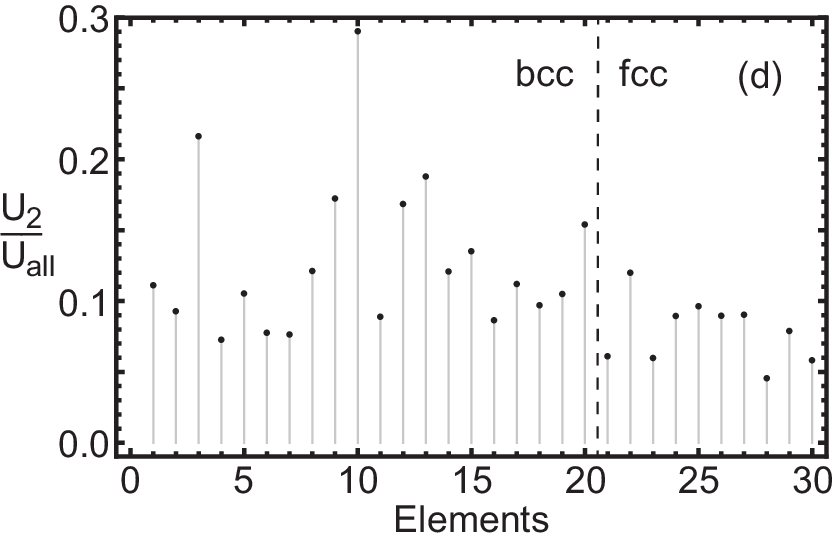}\\
 \caption{(a) Average relative stretching $\frac{\Delta r}{r_m}$ of the bonds between first layer atoms. (b) Share of the energy of stretched bonds between same-layer atoms (summed for the first two layers) in the total surface energy. (c) Relative concentration $n$ of atoms of the first layers, established after relaxation. (d) Contribution of the second layer to the total surface energy.}
 \label{surf-fig6}
\end{figure}

Let us now consider other characteristics of the surface, which correspond to the obtained surface tension coefficients. The values of the average relative stretching $\frac{\Delta r}{r_m}$ of the bonds between the atoms of the first layer are shown in Fig.\ref{surf-fig6}(a). For $fcc$ substances they are in the range $7 \% - 14 \%$. Since the surface tension coefficient shows the energy per unit area, such a stretch significantly affects the value of the surface tension coefficient. When the bonds are stretched by $10 \%$ the area per atom increases by $21 \%$. Both this change in area and the energy of these bonds must be taken into account. At the same time, the obtained stretching of the bonds is relatively small, so that the condition $\frac{\Delta r}{r_m} \ll 1$ for a constant bond rigidity may be considered fulfilled.

The share of the energy of stretched and compressed, but not <<broken>> bonds in the total surface energy is shown in Fig.\ref{surf-fig6}(b). For $fcc$ substances it is in the range of $20 \% - 45 \%$, i.e. it constitutes a significant part of the total surface energy. For $Al$, for example, the surface energy is divided almost equally between the energy of stretched and <<broken>> bonds. Therefore, the surface energy cannot be considered only as the energy of extremely stretched bonds.

The ratio $n$ of the concentrations of atoms in the first two layers is shown in Fig.\ref{surf-fig6}(c). For $fcc$ substances it is in the range of $20 \% - 50 \%$. Due to this difference in concentrations, the number of nearest neighbors for the first layer atoms increases significantly, which is one of the main mechanisms for minimizing the surface energy during relaxation.

The contribution of the energy of the second layer to the total surface energy is shown in Fig.\ref{surf-fig6}(d). For $fcc$ substances it is within $5 \% - 12 \%$, i.e. practically all energy of the surface is concentrated in the first atomic layer. It is clear that its further minimization with the third and subsequent layers will practically have no effect on the resulting surface tension coefficient.

Thus, a simple consideration, but with explicit accounting for the layers stretching and the associated change in the numbers of nearest neighbors, leads to physically reasonable results. The obtained values of the surface tension coefficient are somewhat overestimated for $fcc$ substances. This was to be expected, since only the main mechanisms of surface energy optimization were considered. A more comprehensive consideration may lead to $10-20\%$ corrections. Therefore, in general, it can be considered that the obtained results are in reasonable agreement with the experiment.

\section{Refinement of the coordination numbers.}

The above basic consideration includes a number of assumptions, due to which it becomes possible to calculate all surface characteristics without fitting coefficients. Let us now consider the ways of possible further development of the basic theory, which might improve its accuracy.

The atomic concentrations of the first and second layers are uneven, even without the compression of the second layer. This compression, in turn, changes the coordination number for the atoms of the third layer. The coefficient of the surface tension appears to significantly depend on the exact number of nearest neighbors of the surface atoms. Therefore, to calculate it more precisely, a more accurate relationship is required between the atom's coordination number and the relative concentrations of atoms in the neighboring layers.

Let us consider the coordination numbers for the first three atomic layers of a surface. Let the surface have a fixed area $S$, and let the numbers of atoms in the first three layers be $N_1$, $N_2$, and $N_3$, respectively. The atoms of the first layer have neighbors among the atoms of the same layer $N_{fl}=6$ and among the atoms of the second layer. Let us denote the latter as $C_{12}$. Similarly, for an atom of the second layer, we denote the average numbers of its neighbors from the first and third layers as $C_{21}$ and $C_{23}$, for the third layer, $C_{32}$. For the total number of bonds between layers we can write:

\begin{equation}\label{surf-eq9-1}
\begin{array}{l}
N_1 C_{12} = N_2 C_{21}\\
N_2 C_{23} = N_3 C_{32}\\
\end{array}
\end{equation}

The number of atoms in the layer is related to the average bond stretch as:

\begin{equation}\label{surf-eq9-2}
\begin{array}{l}
N_1 = \frac{S}{\frac{\sqrt{3}}{2} (r_m+\Delta r)^2}\\
N_2 = \frac{S}{\frac{\sqrt{3}}{2} (r_m+\Delta r_1)^2}\\
N_3 = \frac{S}{\frac{\sqrt{3}}{2} (r_m)^2}\\
\end{array}
\end{equation}

The third layer will be considered not stretched, since further such optimization practically will not affect the result. The number of bonds of an atom with atoms from neighboring layers depends on the concentrations of atoms in these layers. In the case of unrelaxed structure, this number should be equal to $N^{no}_s-N_{fl}$. Considering this dependence to be linear for small deviations of the relative concentration from unity:

\begin{equation}\label{surf-eq9-3}
\begin{array}{l}
C_{12} = (N^{no}_s-N_{fl})(1 - b_1 + b_1 \frac{N_2}{N_1})\\
C_{21} = (N^{no}_s-N_{fl})(1 - b_2 + b_2 \frac{N_2}{N_1})=(N^{no}_s-N_{fl})(b_1+(1-b_1)\frac{N_1}{N_2})\\
C_{32} = (N^{no}_s-N_{fl})(1 - b_3 + b_3 \frac{N_2}{N_3})\\
C_{23} = (N^{no}_s-N_{fl})(1 - b_4 + b_4 \frac{N_2}{N_3})=(N^{no}_s-N_{fl})(b_3+(1-b_3)\frac{N_3}{N_2})\\
\end{array}
\end{equation}

It is clear that the coefficients $b_1$ and $b_3$ introduced here must be close to unity. For example, if the number of atoms in the second layer is increased by $10\%$, then the number of interlayer bonds of a first layer atom must also increase by approximately $10\%$. In the previous consideration, a similar coefficient was actually set equal to unity. Here, the coefficients $b_1$ and $b_3$ are introduced explicitly, different for different layers. Generally, they are also different for different substances, since the atomic configuration depends on the interaction potential.

With the new coordination numbers, the expression for the surface tension coefficient will be:

\begin{equation}\label{surf-eq9-4}
\begin{array}{r}
\gamma^{ener} = \frac{N_1 ((N_b-N_{fl}-C_{12}) U_m + \frac{N_{fl}}{2} \frac{1}{2} k \Delta r^2)}{S} + \frac{N_2 ((N_b-N_{fl}-C_{21}-C_{23}) U_m + \frac{N_{fl}}{2} \frac{1}{2} k \Delta r_1^2)}{S}+\frac{N_3 (N_b-N^{no}_s-C_{32}) U_m}{S}\\
\end{array}
\end{equation}

substituting the coefficients Eq.\ref{surf-eq9-3} and minimizing the resulting expression, for the surface tension coefficient we obtain:

\begin{equation}\label{surf-eq9-7}
\begin{array}{l}
\gamma^{ener} =(N_b+2(1-b_3)N_{fl}-(3-2b_3)N^{no}_s)\frac{2U_m}{\sqrt{3} r_m^2} +\\
\\
+ (2-\frac{N_{fl}}{N_{fl}+4(N_b+(2b_1+2b_3-1)N_{fl}-2(b_1+b_3)N^{no}_s)\frac{U_m}{k r_m^2}}-\frac{N_{fl}}{N_{fl}+4((N_b-(2 b_1-1)N_{fl}-2(1-b_1)N^{no}_s)\frac{U_m}{k r_m^2}})\frac{N_{fl} k}{2 \sqrt{3}}\\
\end{array}
\end{equation}

For arbitrary $b_1$ and $b_3$, this expression will not be equal to the $\gamma^{tens}$ obtained using Eq.\ref{surf-eq9}. This is due to the fact that the coefficients $b_1$ and $b_3$ do not correspond to the hexagonal structure of the layers, which was assumed when deriving Eq.\ref{surf-eq9}. Fig.\ref{surf-fig4}(b) shows a comparison of the experimental results with the new surface tension coefficients. Eq.\ref{surf-eq9-7} will give practically the same coefficients as Eq.\ref{surf-eq9} with the values $b_1=b_3=0.85$. The increase to the value $b_1=b_3=0.95$ will make the tension coefficients Eq.\ref{surf-eq9-7} close to the experimental for $fcc$ substances. The individual variation of these coefficients for each substance will easily lead to the ideal fit with the experimental results, but such a number of fitting parameters is obviously senseless.

The coefficients $b_1$ and $b_3$ for each substance should be calculated based on the most energetically favorable configuration of atoms. The energy of the interlayer bonds should also be taken into account. In this way, the dependence of $b_1$ and $b_3$ on $n$ and parameters of the interaction potential may be obtained, which would allow a more accurate calculation of surface characteristics. However, the most energetically favorable structure is apparently not regular at different concentrations of atoms in adjacent layers, which significantly complicates the implementation of such an analysis.

The accuracy of the basic theory may also be improved with accounting for additional effects, in particular associated with the thermal motion of atoms. The kinetic energy can be considered equally distributed between surface and bulk atoms. But it leads to the stretching of all bonds due to the thermal expansion. For a number of substances, it is about $2 \%$ at the melting point. The accounting for this effect will give only a small correction, since the energy of six $2 \%$ stretched bulk bonds is much lower than that of one $12 \%$ stretched bond between surface atoms. Also, the potential energy of oscillations around the mean bond length value may be different for bulk and surface atoms, but this difference is below the order of thermal energy. The thermal energy level $\frac{1}{2} k T_m$ is about ten times lower than the known experimentally surface energy per atom. There is also an energy associated with the average force of attraction of surface atoms to deeper layers of the liquid. Such a force is compensated by the dynamic pressure arising due to the thermal motion. The associated interlayer bond stretching is probably close to that of thermal expansion. If to consider that the thermal scattering is restrained only by the last atomic layer, the associated surface energy will be of the order of thermal energy per surface atom \cite{surf-My}. The potential energy of atoms from the second coordination shell may also be distinguished from that of more distant atoms. The rigidity of the bond may be considered decreasing with distance at $r>r_m$. All there and other factors will provide some minor corrections. But the primary cause of the inaccuracies is the flaws in the description of atomic structure.

\section{Tension of a curved surface.}

The above approaches were developed for the flat surface geometry, but they can be easily extended to curved surfaces. Let us now consider, within the framework of the basic approach, how the coefficient of the surface tension depends on the surface curvature. Let the shape of a surface be a sphere of radius $r$. Then, for the numbers of atoms in the first and second layers, considering that interlayer bonds are unstretched, we can write:

\begin{equation}\label{surf-eq9-8}
\begin{array}{l}
N_1 = \frac{4 \pi r^2}{\frac{\sqrt{3}}{2} (r_m+\Delta r)^2}\\
N_2 = \frac{4 \pi (r-\sqrt{\frac{2}{3}}r_m)^2}{\frac{\sqrt{3}}{2} (r_m+\Delta r_1)^2}\\
n=\frac{N_2}{N_1}=(1-\sqrt{\frac{2}{3}}\frac{r_m}{r})^2 \frac{(r_m+\Delta r)^2}{(r_m+\Delta r_1)^2}
\end{array}
\end{equation}

In this way, the surface geometry changes the dependence of the relative concentration of atoms $n$ on $\Delta r$ and $\Delta r_1$. Any other surface geometry may be considered similarly. Substitution of the relative concentration $n$ from Eq.\ref{surf-eq9-8} into Eq.\ref{surf-eq5} and repetition of the procedure of energy minimization gives the curved surface tension coefficient. Its ratio to the tension coefficient of a flat surface Eq.\ref{surf-eq7} is:

\begin{equation}\label{surf-eq9-9}
\frac{\gamma_r}{\gamma_{\infty}} = 1+\frac{2(\sqrt{6}r-r_m)r_m}{3 r^2} \frac{(N^{no}_s-N_{fl})(N_{fl}+4(N_b-N_{fl})\frac{U_m}{k r_m^2})}{N_{fl}(N_b-N^{no}_s) -8(N_b-N_{fl})(N^{no}_s-N_{fl})\frac{U_m}{k r_m^2}}
\end{equation}

\begin{figure}
 \centering
 \includegraphics[width=8 cm]{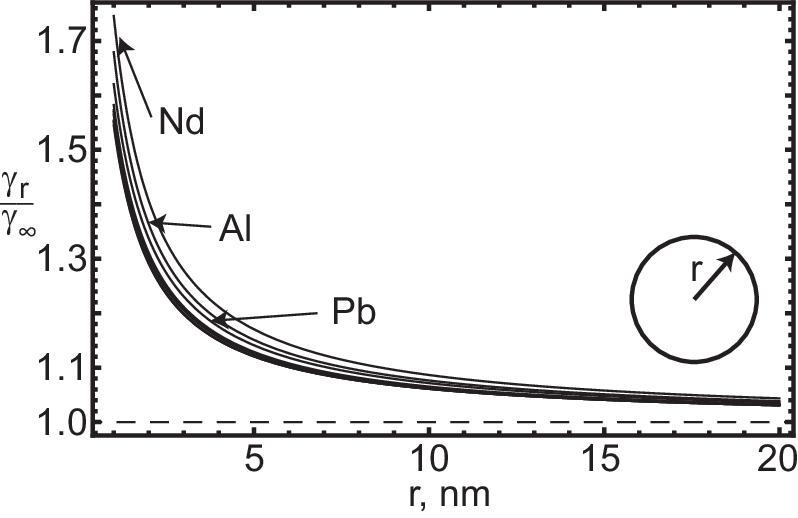}\\
 \caption{Dependence of the surface tension coefficient on the radius of curvature of the surface $r$.}
 \label{surf-fig7}
\end{figure}

Thus, a simple consideration shows that with a decrease in the positive radius of curvature, the surface tension coefficient will grow. This is a consequence of the decrease in the number of nearest neighbors for the first layer atoms. The ratio of the surface tension coefficient to that for a flat surface Eq.\ref{surf-eq9-9} for $fcc$ substances is shown in Fig.\ref{surf-fig7}. The plot starts with a radius of curvature of 1 nm, which corresponds to $50-100$ atoms in the second surface layer. It is visible that the increase by $1.55-1.75$ times is predicted for such a radius of curvature. For $bcc$ substances, the formula Eq.\ref{surf-eq9-9} gives a similar increase by $1.55-2.05$ times. Experimental results on the dependence of the coefficient of surface tension on the radius of curvature are practically absent. But it seems that the obtained result agrees with the results of modeling by molecular dynamics methods.

\section{Discussion and conclusions.}

In the paper, a simple theory of the surface tension of monatomic liquids is proposed. It explicitly accounts for a number of factors that have the most significant effect on the tension coefficient after relaxation. The primary cause for the surface atoms to have a higher energy is a smaller number of their nearest neighbors, compared to the bulk atoms. Because of that, surface atoms have a greater number of extremely stretched or <<broken>> bonds. The tendency to minimize their number and distribute the potential energy more evenly between atoms bonds results in the appearance of the average stretch of bonds between first layer atoms. The potential energy of these bonds significantly contributes to the surface energy. Difference in concentrations of atoms in the top layers lead to the changes in their coordination numbers, compared to the unrelaxed atomic structure. That significantly affects the value of the surface tension coefficient. These are the primary factors that determine the energetics of the surface after relaxation.

Using this basic approach, we have calculated the main characteristics of the surface. The number of <<broken>> bonds of surface atoms after relaxation was obtained. Their energy makes the main contribution to the surface energy. The stretching of the bonds between atoms of the first layer and their energy was also obtained. It was shown that the calculated tension of a unit surface length coincides with the energy of a unit surface area. The contribution of the second atomic layer was accounted for. It was shown that the compression of the second layer leads to the further significant optimization of the surface energy. In the result, the expressions for the surface tension coefficient were obtained, for both flat and curved surfaces.

For comparison with experimental results, the parameters of the pair interaction potential of atoms were determined, for simple monoatomic substances. For this purpose, the heat of vaporization, the coefficient of volumetric compressibility, and the molar volume of substance were used, all at the melting temperature. Since the experimental coefficients of compressibility and surface tension are known with an accuracy of about $10 \%$, a range of possible values of the ratio of theoretical and experimental tension coefficients was calculated. For all ten $fcc$ substances, these ranges fall within the $1.0-1.3$ interval. Thus, the obtained theoretical values appeared to be overestimated. One of the reasons for that is that only the main mechanisms of surface energy optimization were considered. The contribution of energy of the stretched bonds within the top atomic layers was $20 \% - 45 \%$, with an average stretching of $7 \% - 14 \%$ of the first atomic layer. Also, for reference, data for twenty $bcc$ substances were substituted into the formulas for $fcc$ substances. All the ratios of theoretical and experimental surface tension coefficients are within the $0.65-1.35$ range. Thus, the theoretically obtained values of the surface tension coefficients reasonably coincided with the experimental values.

\section{Appendix 1.}

\begin{figure}
 \centering
 \includegraphics[width=4 cm]{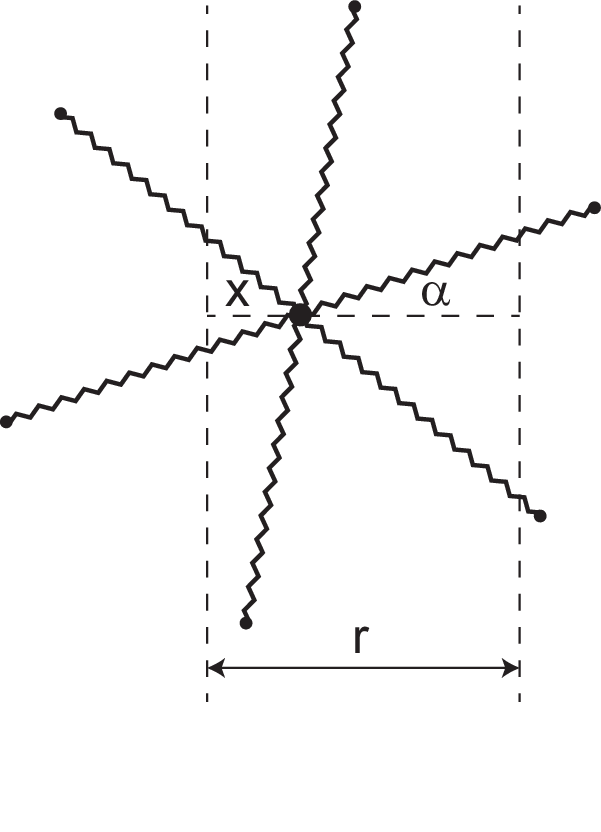}
 \includegraphics[width=6 cm]{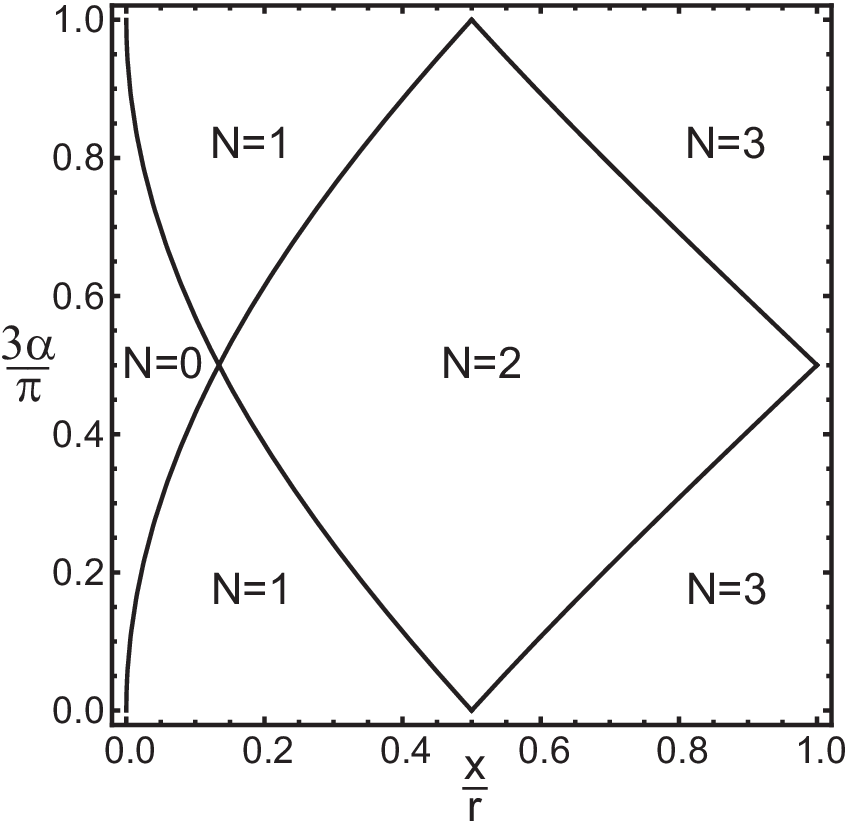}\\
 \caption{(a) An atom of the surface of a simple substance and $N_{fl}=6$ neighboring atoms at a distance $\overline{r}=r_m+\overline{\Delta r}$ from it. The selected atom has a random coordinate $x \in [0,\overline{r}]$ and a random angle $\alpha \in [0,\frac{\pi}{3}]$ of direction to the first neighboring atom. (b) The number of bonds of the selected atom that intersect the right boundary of the strip of width $\overline{r}$ in dependence on $x$ and $\alpha$.}
 \label{surf-app-fig1}
\end{figure}

Let us consider how the tension of a liquid surface is related to the average stretching of bonds between neighboring atoms of the top surface layer. We will consider that the surface structure is characterized by short-range order, but it has no long-range order. Due to the short-range order, each atom of the first surface layer has $N_{fl}=6$ nearest neighbors among the atoms of this layer. All surface atoms are identical, the average distance between them will be equal to $\overline{r}=r_m+\overline{\Delta r}$. Thus, all bonds between surface atoms are on average stretched by $\overline{\Delta r}$. If we draw an imaginary line on the surface, then all the forces $k \overline{\Delta r}$ intersecting it will contribute to the surface tension force. If a bond of an atom intersects the line, then this atom is at a distance of no more than $\overline{r}$ from it. Therefore, we select a strip of width $\overline{r}$ on the surface and calculate the average contribution made by one atom of this strip. Let the selected atom have a coordinate $x \in [0,\overline{r}]$, and let one of its bonds have the angle $\alpha$ that falls within the range $\alpha \in [0,\frac{\pi}{3}]$ (see Fig.\ref{surf-app-fig1}(a)). If this bond intersects the right boundary of the strip, then it makes a contribution $k \overline{\Delta r} \cos \alpha$ to the average surface tension force. The values $x$ and $\alpha$ will be considered random and distributed uniformly within their range. Depending on these values, the band boundary may be crossed by zero to three bonds of the atom. The dependence of the number of crossings on $x$ and $\alpha$ is shown in Fig.\ref{surf-app-fig1}(b). If the band boundary is crossed by one bond, its contribution will be:

\begin{equation}\label{surf-app-eq1}
\overline{f^{N=1}} = 2 \frac{3}{\pi \overline{r} } k \overline{\Delta r} \,  (\underset{x=0}{\stackrel{\frac{2-\sqrt{3}}{2}\overline{r}}\int} \,\,\, \underset{\alpha=0}{\stackrel{\arccos(1-\frac{x}{\overline{r}})}\int} \!\!\!\!\!\!\!\!\! \cos \alpha \,\,\, dx d\alpha \,\, + \underset{x=\frac{2-\sqrt{3} }{2}\overline{r}}{\stackrel{\frac{\overline{r}}{2}}\int} \!\!\!\! \underset{\alpha=0}{\stackrel{\frac{\pi}{3}-\arccos(1-\frac{x}{\overline{r}})}\int} \!\!\!\!\!\!\!\!\!\!\!\! \cos \alpha \,\,\, dx d\alpha) = \frac{1}{4}  k \overline{\Delta r}
\end{equation}

The coefficient $\frac{3}{\overline{r} \pi}$ is a normalization coefficient, since the probability that the variables $x$ and $\alpha$ fall within the ranges $[0,\frac{\pi}{3}]$ and $[0,\overline{r}]$ must be equal to unity. Similarly, for two and three bonds crossing the boundary of the strip, we obtain:

\begin{equation}\label{surf-app-eq2}
\overline{f^{N=2}} = \frac{3 k \overline{\Delta r}}{\pi \overline{r} }  ( \!\!\!\!\!\!\!\! \underset{x=\frac{2-\sqrt{3}}{2}\overline{r}}{\stackrel{\frac{\overline{r}}{2}}\int} \!\,\,\, \underset{ \alpha=\frac{\pi}{3}-\arccos(1-\frac{x}{\overline{r}})}{\stackrel{\arccos(1-\frac{x}{\overline{r}})}\int} \!\!\!\!\!\!\!\!\!\!\!\!\!\!\!\!\! (\cos \alpha + \cos (\alpha-\frac{\pi}{3})) \, dx d\alpha \, + \!\!\!\! \underset{x=\frac{\overline{r}}{2}}{\stackrel{\overline{r}}\int} \!\,\,\, \underset{\alpha=\arccos(1-\frac{x}{\overline{r}})-\frac{\pi}{3}}{\stackrel{\frac{2\pi}{3}-\arccos(1-\frac{x}{\overline{r}})}\int} \!\!\!\!\!\!\!\!\!\!\!\!\!\!\!\!\! (\cos \alpha + \cos (\alpha-\frac{\pi}{3})) \, dx d\alpha) = \frac{3}{4}  k \overline{\Delta r}
\end{equation}

and

\begin{equation}\label{surf-app-eq3}
\overline{f^{N=3}} = 2 \frac{3}{\pi \overline{r} } k \overline{\Delta r} \, ( \!\! \underset{x=\frac{\overline{r}}{2}}{\stackrel{\overline{r}}\int} \!\,\,\, \underset{\alpha=0}{\stackrel{\arccos(1-\frac{x}{\overline{r}})-\frac{\pi}{3}}\int} \!\!\!\!\!\!\!\!\!\!\! (\cos \alpha + \cos (\alpha-\frac{\pi}{3}) + \cos (\alpha+\frac{\pi}{3})) \, dx d\alpha = \frac{1}{2}  k \overline{\Delta r}
\end{equation}

Thus, one atom of the selected strip makes a whole contribution of $\frac{3}{2} k \overline{\Delta r}$ to the average surface tension force. The area per one atom of a surface with hexagonal structure is equal to $\frac{\sqrt{3}}{2} \overline{r}^2$, hence the number of atoms in a strip of width $\overline{r}$ and length $l$ is $N=\frac{\overline{r} l}{\frac{\sqrt{3}}{2} \overline{r}^2}$. Therefore, the strip length per one atom is $\frac{\sqrt{3}}{2} \overline{r}$. These coefficients were used in the formula of Eq.\ref{surf-eq9} and further on.

Similarly, one can consider a bulk atom with its first coordination shell and calculate how the bulk pressure is related to the average stretch of the bonds between the atoms. Under small volumetric compression by the value $\Delta V$, due to the application of a uniform pressure $P$, all linear distances between atoms will be reduced proportionally to $\frac{\Delta V}{3 V}$. The average distance between nearest neighbors $\overline{r}$ will be reduced by the value $\Delta \overline{r} = \overline{r} \frac{\Delta V}{3 V}$. The coefficient of volume compression by definition is equal to $k_T=\frac{1}{P}\frac{\Delta V}{V} \approx \frac{3}{P}\frac{\Delta \overline{r}}{\overline{r}}$. In this case, additional forces $\Delta f_i$ arise between atoms, compensating for the applied pressure. At small pressure applied, we can assume that $<\Delta f_i>=k \Delta \overline{r}$, where $k$ is the required bond stiffness. The external pressure $P$ is related to the forces $\Delta f_i$ as:

\begin{equation}\label{surf-app-eq4}
P = \frac{F}{S} = \frac{\sum_i \Delta f_i \cos \alpha_i}{S}
\end{equation}

The average number of bonds intersecting a plane of area $S$ and their contribution to the average pressure force can be calculated similarly to what was done above for tension force. Since the integrals arising from averaging are rather cumbersome, we will present only the result. In the case of the $fcc$ structure, the average contribution to the pressure force of one atom located at a distance less than $\overline{r}$ from a plane is $2 k \Delta \overline{r}$. This can also be verified numerically using the Monte Carlo method. Then for the relationship of the pressure and compression of bonds we have:

\begin{equation}\label{surf-app-eq5}
P = \frac{2 k \Delta \overline{r}}{S_0} = 2 k \Delta \overline{r} \frac{k_{pac} \overline{r}}{\frac{1}{6} \pi \overline{r}^3} = \frac{12 k_{pac}}{\pi} \frac{1}{\overline{r}} \frac{\Delta \overline{r}}{\overline{r}} k = \frac{12 k_{pac}}{\pi} \frac{1}{\overline{r}} \frac{P k_T}{3} k
\end{equation}

from where, considering the bonds of bulk atoms to be practically unstretched $\overline{r}=r_m$, we obtain Eq.\ref{surf-eq12}:

\begin{equation}\label{surf-app-eq6}
k=\frac{\pi}{4 k_{pac}} \frac{r_m}{k_T}
\end{equation}

\section{Appendix 2. Table of the constants used.}

\begin{table}
  \centering
  \begin{tabular}{|c|c|c|c|c|c|c|}
    \hline
     & Melting  & Surface & Isothermal & Molar & Coord.  \\
    Substance & Temperature  & Tension & compressibility & volume $V_m$,  & number\\
      & $T_m$, K & $\gamma$, $\mathrm{\frac{mN}{m}}$ & $k_T$ , 1/GPa & $10^{-29} m^3$/atom & $N_b$\\
     \hline
   Ag & 1234 & 882-964 & 0.0176-0.0186 & 1.92 & 12 \\
   Al & 933 & 865-930 & 0.0242-0.0263 & 1.88 & 12 \\
   Au & 1338 & 1105-1185 & 0.0131 & 1.89 & 12 \\
   Co & 1768 & 1790-1885 & 0.0097-0.0107 & 2.56 & 12 \\
   Cu & 1358 & 1274-1393 & 0.0136-0.0151 & 2.57 & 12 \\
   Nd & 1294 & 685-689 & 0.0341-0.0455 & 3.58 & 12 \\
   Ni & 1728 & 1723-1854 & 0.0098-0.0108 & 1.23 & 12 \\
   Pb & 601 & 442-480 & 0.0341-0.035 & 3.22 & 12 \\
   Pd & 1828 & 1475-1500 & 0.0128-0.0132 & 1.68 & 12 \\
   Pt & 2042 & 1707-1865 & 0.0084 & 1.71 & 12 \\
   \hline
   Ba & 1000 & 226-267 & 0.178-0.178 & 6.87 & 14\\
   Ca & 1115 & 337-366 & 0.094-0.110 & 4.88 & 14\\
   Ce & 1071 & 707-794 & 0.0464-0.047 & 3.48 & 14\\
   Cr & 2180 & 1628-1700 & 0.010-0.0122 & 1.37 & 14\\
   Cs  & 302 & 69-86 & 0.63-0.688 & 12.0 & 14\\
   Fe & 1811 & 1780-1918 & 0.0104-0.0118 & 1.32 & 14\\
   Gd & 1585 & 664-810 & 0.025-0.042 & 3.35 & 14\\
   K & 337 & 94-123 & 0.382-0.402 & 7.86 & 14\\
   La & 1191 & 701-745 & 0.039-0.0552 & 3.87 & 14\\
   Li  & 454 & 398-399 & 0.093-0.11 & 2.22 & 14\\
   Mn & 1519 & 1100-1219 & 0.0173-0.0317 & 1.58 & 14\\
   Na & 371 & 191-206 & 0.186-0.210 & 4.12 & 14\\
   Pr & 1205 & 690-743 & 0.0523-0.0537 & 3.54 & 14\\
   Rb  & 312 & 85-90 & 0.474-0.493 & 9.58 & 14\\
   Sc & 1814 & 870-939 & 0.0325-0.0364 & 2.79 & 14\\
   Sr & 1050 & 286-289 & 0.131-0.137 & 6.15 & 14\\
   Ta & 3290 & 2016-2150 & 0.01 & 2.01 & 14\\
   Ti & 1941 & 1390-1650 & 0.014-0.0163 & 1.93 & 14\\
   V & 2183 & 1849-1950 & 0.011-0.0143 & 1.58 & 14\\
   W & 3695 & 2220-2316 & 0.0095 & 1.73 & 14\\
   \hline


  \end{tabular}
  \caption{Table of the used constants and the ranges of their experimental values for simple substances in the liquid state at the melting temperature. The surface tension coefficient data is from the survey \cite{surf-Keen}, and the compressibility data is from the survey \cite{surf-Marc}.}\label{eqstat-ap-tab1}
\end{table}


\begin{thebibliography}{99}

\bibitem{surf-Zhao} M. Zhao, W. Zheng, J. Li, and Z. Wen. M. Gu and C. Q. Sun. Atomistic origin, temperature dependence, and responsibilities of surface energetics: An extended broken-bond rule. PHYSICAL REVIEW B 75, 085427 (2007)
    DOI: 10.1103/PhysRevB.75.085427

\bibitem{surf-Bru} N. Bruot and F. Caupin. Curvature Dependence of the Liquid-Vapor Surface Tension beyond the Tolman Approximation. PRL \textbf{116}, 056102 (2016)
    DOI: 10.1103/PhysRevLett.116.056102

\bibitem{surf-Wang} D. Wang, Z. Hu, G. Peng, and Y. Yin. Surface Energy of Curved Surface Based on Lennard-Jones Potential. Nanomaterials \textbf{11}, 686 (2021)
    https://doi.org/10.3390/nano11030686

\bibitem{surf-Kim} D. Kim, J. Kim, J. Hwang, D. Shin, S. Ana, and W. Jhe. Direct measurement of curvature-dependent surface tension of an alcohol nanomeniscus. Nanoscale \textbf{13}, 6991 (2021)
    DOI: 10.1039/d0nr08787d

\bibitem{surf-Duan} H. Duan, Z. Cui, Y. Xue, Q. Fu, X. Chen and R. Zhang. Determination method and size dependence of interfacial tension between nanoparticles and a solution. Langmuir \textbf{34}(30), 8792-8797 (2018)
DOI: 10.1021/acs.langmuir.8b01702

\bibitem{surf-Hole} D. Holec, P. Dumitraschkewitz, D. Vollath and F. D. Fischer. Surface Energy of Au Nanoparticles Depending on Their Size and Shape. Nanomaterials \textbf{10}, 484(15) (2020)
    doi:10.3390/nano10030484

\bibitem{surf-Lenn} J. E. Lennard-Jones and J. Corner. The calculation of surface tension from intermolecular forces. Trans. Faraday Soc. \textbf{36}, 1156-1162, (1940)

\bibitem{surf-Dur} M. Durand. Mechanical approach to surface tension and capillary phenomena. American Journal of Physics \textbf{89}(3), 261-266 (2021)

\bibitem{surf-Berr} M. V. Berry. The molecular mechanism of surface tension. Physics Education \textbf{6}(2), 79-84 (1971)

\bibitem{surf-Hass} S. M. Hassanizadeh. The Origin of Surface Tension. InterPore Journal, 1(1), ipj260424–3 (2024)
https://doi.org/10.69631/ipj.v1i1nr21

\bibitem{surf-Sop} M. Sophocleous. Understanding and explaining surface tension and capillarity: an introduction to fundamental physics for water professionals. Hydrogeology Journal (2010) 18: 811–821
    DOI 10.1007/s10040-009-0565-5

\bibitem{surf-Dav} H. T. Davis and L. E. Scrlven. A Simple Theory of Surface Tension at Low Vapor Pressure. The Journal of Physical Chemistry \textbf{80(25)}, 2805-2806 (1976)

\bibitem{surf-Mona} L. K. Antanovskii. Microscale theory of surface tension. Phys. Rev. E \textbf{54}(6), 6285-6290 (1996)

\bibitem{surf-Weir} J. Weir. Implications from the ratio of surface tension to bulk modulus and nearest neighbour distance, for planar surfaces. Proc. R. Soc. A  \textbf{464}, 2281–2292 (2008)
    doi:10.1098/rspa.2007.0360

\bibitem{surf-Was} Y. Waseda and K. T. Jacob. Refinement of the Correlation between Isothermal Compressibility and Surface Tension of Liquid Metals. phys. stat. sol. (a) 68, K l l 7 (1981)

\bibitem{surf-Iida} T. Iida and R. Guthrie. Performance of a Modified Schytil Model for the Surface Tension of Liquid Metallic Elements at Their Melting Point Temperatures. METALLURGICAL AND MATERIALS TRANSACTIONS B \textbf{41}(B), 437-447 (2010)
DOI: 10.1007/s11663-009-9330-3

\bibitem{surf-Bla} S. Blairs. Correlation between surface tension, density, and sound velocity of liquid metals. Journal of Colloid and Interface Science \textbf{302}, 312–314 (2006)
    doi:10.1016/j.jcis.2006.06.025

\bibitem{surf-Schy} F. Schytil. Z. Naturforsh. \textbf{4}, pp. 191–194 (1949)

\bibitem{surf-My} D. M. Naplekov and V. V. Yanovsky. Equation of state of a small system with surface degrees of freedom. arxiv

\bibitem{surf-Chen} J. Chen, S. Xu, B. Wang, X. Fan, D. J. Singh, and W. Zheng. Insights into the surface tension and superficial density peak of molten metals from molecular dynamics. Acta Materialia, \textbf{276}, 120149 (2024)

\bibitem{surf-Nogi} K. Nogi, K. Ogino, A. Mclean, and W. A. Miller. The Temperature Coefficient of the Surface Tension of Pure Liquid Metals. Metallurgical Transactions B \textbf{17}(B), p.163-170 (1986)

\bibitem{surf-Egam} T. Egami and C. W. Ryu. World beyond the nearest neighbors. J. Phys.: Condens. Matter \textbf{35}, 174002(6) (2023)
https://doi.org/10.1088/1361-648X/acbe24

\bibitem{surf-Keen} B. J. Keene. Review of data for the surface tension of pure metals. International Materials Reviews \textbf{38}(4), 157-192, (1993)

\bibitem{surf-Marc} Y. Marcus. On the Compressibility of Liquid Metals. J. Chem. Thermodynamics (2016),
doi: http://dx.doi.org/10.1016/j.jct.2016.07.027

\end{thebibliography}
\end{document}